# EPISTEMIC SUBORDINATION: GENERATIVE AI AND THE INFRASTRUCTURE OF KNOWLEDGE

*Gilad Abiri and Emanuel V. Towfigh*[*]

## INTRODUCTION

In this essay, we argue that generative AI does something no previous technology has done: it encodes the majority's way of knowing as the infrastructure of knowledge itself. We call this epistemic subordination. The harm is not bias in the conventional sense, though biased decisions are one of its consequences. It is the construction of a statistical knowledge system in which one epistemology is the default and every other is a deviation. The concept of epistemic subordination unifies what the current legal conversation treats as separate problems: discriminatory AI outputs,[1] threats to minority cultures,[2] and the narrowing of democratic discourse.[3] We show that these are three manifestations of a single architectural feature of generative AI, and that existing legal frameworks in each domain are structurally unable to reach it.

The mechanism is architectural, and it unfolds in three stages. First, most models are trained on data from the internet. The internet is many things, but it is not epistemically neutral: it is dominated by the languages, assumptions, and cultural frameworks of the global majority.[4] That dominance becomes the model's default. Second, the training process makes that default irreversible. Data is compressed into billions of statistical parameters that cannot be unpacked, isolated, or traced to their source.[5] Third, value alignment techniques — the tools designed to make AI safe and fair — adjust what the model says without altering what it knows.[6] It follows that epistemic subordination cannot be addressed at the level of outputs, applications, or individual decisions. The harm is produced in the construction of the model, and it is therefore at the level of construction that law must intervene. This is the central normative claim of the essay.[7] Current legal frameworks — anti-discrimination law, cultural and linguistic rights, viewpoint pluralism doctrines, and even AI-specific regulation — all operate downstream. They govern what AI systems do, not how they are built. A legal response adequate to epistemic subordination would need to reach the training process itself: the composition of data, the methods of alignment, and the architecture of the model.

---

[1] See *infra* Part II.A.
[2] See *infra* Part II.B.
[3] See *infra* Part II.C.
[4] See Alice Xiang, *Fairness & Privacy in an Age of Generative AI*, 25 *Colum. Sci. & Tech. L. Rev.* 288, 290–298 (2024).

The essay proceeds as follows. Part I examines the training process that produces epistemic subordination. Part II shows how the resulting harm cuts across anti-discrimination law, cultural and linguistic rights, and democratic viewpoint pluralism, and why each legal domain fails to address it. Part III argues that regulation must be redirected from the outputs of generative AI to the infrastructure of its creation, and surveys emerging technical approaches that demonstrate the feasibility of such intervention.

## I. THE ARCHITECTURE OF EPISTEMIC SUBORDINATION

---

[5] Simon Chesterman, *Through a Glass Darkly: Artificial Intelligence and the Problem of Opacity*, 69 *Am. J. Compar. L.* 271, 278 (2021) (explaining that deep learning methods are opaque by design, relying on machine learning rather than transparent decision trees).
[6] See Jianfeng Cao, *The Role of AI Value Alignment in China's Approach to AI Ethics and Governance*, *German L.J.* (forthcoming 2026) (Manuscript at III.C).
[7] See *infra* Part III.

Generative AI systems differ from their predictive predecessors in a way that is fundamental to the problem of discrimination. Predictive AI models (also called ADM systems) are trained on structured, labeled datasets compiled for specific tasks: loan defaults, recidivism rates, hiring outcomes.[8] Their biases, while serious, are artifacts of identifiable design choices — which variables to include, which proxies to permit, which historical patterns to encode.[9] Generative AI, by contrast, is trained on the unstructured totality of human expression available on the internet: websites, books, forums, social media, code repositories, news archives.[10] This training corpus is not curated for any particular task. It is a cultural corpus, and its composition reflects the demographics of digital publication: overwhelmingly English-language, Western, and produced by populations with the education and access to publish online.[11] The result is a model whose foundational parameters encode the linguistic patterns, cultural assumptions, and normative frameworks of the majority as the statistical default.

The biased baseline created by training data is made irreversible by the opacity of the model itself. During training, data is broken into statistical relationships across and between billions of parameters. The result is a model whose outputs cannot be practically traced back to any identifiable source in the training corpus. This opacity is inherent and structural. As Simon Chesterman explains, "[s]ome deep learning methods are opaque effectively by design, in the sense that they rely on machine learning rather than rules that can be articulated through a transparent decision tree."[12] For anti-discrimination law, this means there is no lever to pull in order to mitigate the model's bias, no variable to isolate, no input to remove, and no causal chain to reconstruct.[13]

---

[8] Indra Spiecker & Emanuel V. Towfigh, *Coded Bias, The General Equal Treatment Act and Protection Against Discrimination through Algorithmic Decision-Making Systems,* 21 (2023).

[9] Dai Xin, Regulatory Alternatives to the AI Social Scoring Ban: A Comparative Perspective, *German L.J.* (forthcoming 2026) (Manuscript at III.B); Danielle Keats Citron & Frank Pasquale, *The Scored Society: Due Process for Automated Predictions,* 89 Wash. L. Rev. 1, 4 (2014).

[10] Tom B. Brown et al., Language Models Are Few-Shot Learners 8–9 tbl.2.2 (OpenAI, Working Paper, 2020), https://arxiv.org/abs/2005.14165 (detailing GPT-3's training corpus of approximately 499 billion tokens drawn from filtered Common Crawl web data, WebText2, two book corpora, and Wikipedia).

[11] Emily M. Bender et al., On the Dangers of Stochastic Parrots: Can Language Models Be Too Big?, Proc. 2021 ACM Conf. on Fairness, Accountability & Transparency 610, 613–15 (2021), https://dl.acm.org/doi/10.1145/3442188.3445922 (arguing that internet-sourced training data systematically overrepresents hegemonic viewpoints — predominantly English-language, Western, young, and from populations with the access and education to publish online).

The techniques developed to align AI systems with human values operate at several stages but share a common limitation. At the fine-tuning stage, reinforcement learning from human feedback (RLHF) adjusts model outputs based on human evaluators' preferences,[14] while constitutional AI embeds normative principles directly into the training process.[15] At deployment, system-level instructions constrain the model's behavior, and content moderation algorithms filter its outputs.[16] Each of these techniques adjusts what the model produces. None restructures the epistemic foundation on which it operates.[17] When value alignment does attempt to address cultural bias directly, the results reveal the depth of the problem: Google's Gemini, instructed to generate diverse images, produced historically absurd results — racially diverse Nazi soldiers, for instance — because the model's baseline had no framework for contextual diversity, only a statistical default that could be crudely overridden.[18]

---

[12] Simon Chesterman, *Through a Glass Darkly: Artificial Intelligence and the Problem of Opacity*, 69 *Am. J. Compar. L.* 271, 278 (2021).
[13] See Emanuel V. Towfigh, *Generative AI and Anti-Discrimination Law*, *German L.J.* (forthcoming 2026) (manuscript at C.IV.1.).
[14]Long Ouyang et al., *Training Language Models to Follow Instructions with Human Feedback*, 36th Conf. on Neural Info. Processing Sys. (2022).
[15] Gilad Abiri, *Public Constitutional AI*, 59 *Ga. L. Rev.* 601 (2025) (proposing participatory alternatives to corporate-defined AI constitutional principles).
[16] *See* OpenAI, GPT-4 Technical Report 53–67 (2023), https://arxiv.org/abs/2303.08774 (describing system-level instructions that constrain model behavior and a model-assisted safety pipeline that filters outputs at deployment).
[17] See *Towfigh*, *supra* note 13, at C.IV.2.

Generative AI encodes majority-culture epistemology as its statistical default. In a system in which AI is a primary source of information and culture, every other way of knowing, reasoning, and communicating is not merely underrepresented but structurally subordinated, measured against a baseline it had no part in setting. The challenge of epistemic subordination is not only that generative AI will produce biased decisions, but that entire forms of knowledge and culture are subordinated. The implications include issues of classical anti-discrimination law but reach further — to constitutional protections of culture, language, and religion, and to the epistemic foundations of democratic deliberation.

## II. The Legal Impact of Epistemic Subordination

Anti-discrimination law, language and cultural rights, and religious freedom operate through different doctrines and ideas, but they share a common general purpose: the protection of minorities from majority dominance. Part I has shown that generative AI disrupts these mechanisms in different ways. The following sections trace the impact of epistemic subordination across three legal domains, showing in each case how existing doctrine fails to reach the harm.

### *A. Anti-Discrimination: The Invisible Baseline*

---

[18] Thao Phan, *Black Nazis, Asian Vikings: White Paranoia Haunts Generative AI*, Austl. Acad. of the Humans. (Apr. 2024), https://humanities.org.au/power-of-the-humanities/black-nazis-asian-vikings-and-other-problems-with-generative-ai/.

The empirical evidence of generative AI's discriminatory outputs is substantial. Studies show that large language models associate speakers of African American English with negative stereotypes, generating descriptors such as "dirty," "lazy," and "aggressive" in response to the dialect itself.[19] Medical AI systems evaluate LGBTQIA+ patients for mental health concerns at six to seven times the clinical baseline and recommend less aggressive diagnostic imaging for lower-income patients.[20] Text-to-image generators amplify demographic stereotypes at scale, associating high-status professions with lighter skin and criminality with darker skin.[21]

These are not isolated malfunctions. They are the baseline of the model expressing itself. This challenges the basic logic of anti-discrimination law, which operates by contrast: it identifies discrimination by comparing a challenged practice against a neutral baseline. A bank that rejects loan applicants from a particular neighborhood, a university that weights standardized test scores known to correlate with race — in each case, the law measures the practice against what a non-discriminatory process would look like. Disparate impact doctrine, rooted in Title VII[22] and *Griggs v. Duke Power Co.*,[23] extends this logic: even absent discriminatory intent, a practice that produces disproportionate effects on a protected group can be struck down. European anti-discrimination law, grounded in Article

[19] Valentin Hofmann et al., *AI Generates Covertly Racist Decisions About People Based on Their Dialect*, 633 *Nature* 147 (2024).

[20] Mahmud Omar et al., *Sociodemographic Biases in Medical Decision Making by Large Language Models*, 31 *Nature Med.* 1873 (2025).

[21] Federico Bianchi et al., *Easily Accessible Text-to-Image Generation Amplifies Demographic Stereotypes at Large Scale*, 2023 ACM Conf. on Fairness, Accountability & Transparency 1493.

[22] Civil Rights Act of 1964, tit. VII, 42 U.S.C. §§ 2000e to 2000e-17.

[23] Griggs v. Duke Power Co., 401 U.S. 424 (1971).

14 of the European Convention on Human Rights[24] and Article 21 of the EU Charter of Fundamental Rights,[25] follows a parallel structure.

Applied to AI, this framework initially appears to hold promise. If an algorithm produces outcomes that disproportionately disadvantage a protected group, the statistical disparity alone can trigger scrutiny, regardless of whether the developer intended any harm. But disparate impact still requires an identifiable practice that produces the disparity. A diploma requirement, a credit threshold, a variable in a sentencing algorithm — each can be isolated, tested, and if necessary, struck down. Generative AI offers no equivalent target. The discriminatory output is not the product of any discrete criterion but of the entire architecture.

As Towfigh has argued, anti-discrimination law was built for a world of individual misconduct by identifiable actors.[26] It presupposes that discrimination can be traced to a specific decision, made by a specific entity, using identifiable criteria. Generative AI dissolves all three: the "decision" is a probabilistic output, the "entity" is a model trained by one company, fine-tuned by another, and deployed by a third, and the "criteria" are billions of statistical parameters that cannot be individually examined. The harm is no longer confined to allocative effects; it reaches beyond the discriminatory distribution of opportunities or resources tied to protected attributes. The harm is the result of inaccurate or stereotypical representations.[27] The doctrinal tools that anti-discrimination law has developed over decades — intent, causation, disparate impact — lose their grip.[28]

---

[24] Convention for the Protection of Human Rights and Fundamental Freedoms art. 14, Nov. 4, 1950, E.T.S. No. 5.
[25] Charter of Fundamental Rights of the European Union art. 21.
[26] See *Towfigh*, *supra* note 13, at C.I (arguing that anti-discrimination law is built for a world of individual misconduct by identifiable actors).

*B. Cultural, Linguistic, and Religious Rights: The Threat from Within*

The rights examined in this section — cultural, linguistic, and religious — are doctrinally distinct but functionally unified. As Will Kymlicka observed, the state "unavoidably promotes certain cultural identities, and thereby disadvantages others."[29] Both society at large and the state inevitably operate in the dominant language, enforce formal and informal norms drawn from the majority's culture, and shape public life around assumptions that are treated as universal but are not. Many understand linguistic, cultural and religious rights as an attempt to counteract this structural asymmetry.[30] In other words, these are rights that seek to protect minority cultures. In U.S. law, the Free Exercise Clause and the broader First Amendment tradition shield minority religious practice, expression, and cultural life from majority imposition.[31] European law is more explicit: EU Charter Article 22 enshrines respect for cultural, religious, and linguistic diversity,[32] while Article 10 protects freedom of thought, conscience, and religion.[33] Canadian constitutional law, the context from which Kymlicka's theory emerged, goes furthest: the Charter of Rights

[27] see *Xiang*, *supra* note 4, at 291.
[28] See *Towfigh*, *supra* note 13 at C.II–C.III.
[29] Will Kymlicka, *Multicultural Citizenship: A Liberal Theory of Minority Rights* 108 (1995).
[30] See *Kymlicka*, *supra* note 26, at 76 (defining societal culture as one that provides its members with meaningful ways of life across the full range of human activities).
[31] *See e.g.,* W. Va. State Bd. of Educ. v. Barnette, 319 U.S. 624, 638 (1943) ("The very purpose of a Bill of Rights was to withdraw certain subjects from the vicissitudes of political controversy, to place them beyond the reach of majorities.").
[32] Charter of Fundamental Rights of the European Union art. 22.
[33] Charter of Fundamental Rights of the European Union art. 10, ¶ 1.

and Freedoms commits to preserving and enhancing the multicultural heritage of Canadians and protects official bilingualism as a structural feature of the state.[34]

These rights operate through a common mechanism: they carve out institutional space in which minority cultures can function on their own terms, exempt from the norms and assumptions of the majority. Bilingual education programs allow minority-language communities to transmit their language and the knowledge embedded in it to the next generation. Religious exemptions permit minority faiths to follow practices that majority norms would otherwise prohibit — from ritual dress codes to dietary laws to Sabbath observance. Cultural rights protections enable indigenous and minority communities to maintain institutions that operate according to their own traditions. The underlying logic is the same: left unprotected, minority cultures will not survive contact with the majority's institutional and cultural dominance.

Generative AI poses a distinct threat to this protective structure, one that previous knowledge technologies did not. Books, websites, and digital platforms carry cultural content, but that content can be selected. A minority school can curate its own library, choose its own curriculum, host its own online resources. The cultural orientation of these technologies resides in the content, not in the medium. Generative AI reverses this. The cultural bias is not in the content but in the architecture of the model itself. And unlike a book or a website, a foundation model cannot be produced by a minority community. Training one costs hundreds of millions of dollars and requires massive amounts of text and data that simply do not exist in the languages and cultures of most minorities, certainly not at the scale the internet provides for dominant cultures.[35] Recent multilingual benchmarking confirms the disparity: the best-performing large language models achieve approximately seventy percent accuracy in English but only forty percent in languages such as Swahili, a gap that widens further for smaller models.[36] The construction of knowledge infrastructure is permanently beyond the reach of the communities most affected by its defaults. The consequence is that generative AI penetrates the institutional spaces these rights were designed to protect. A religious school that uses an AI-assisted curriculum imports, along with the technology, a normative framework that defaults to secular-Western reasoning.[37] A minority-language classroom that relies on AI for research or translation imports a model whose baseline treats that language as marginal. Cultural, linguistic, and religious protections were built to shield minority institutions from external pressure. A technology that restructures knowledge from within does not render these protections obsolete, but it introduces a challenge they were never designed to meet.

[34] Canadian Charter of Rights and Freedoms, Part I of the Constitution Act, 1982, being Schedule B to the Canada Act, 1982, c 11, § 27.

---

[35] Isabelle A. Zaugg, Anushah Hossain & Brendan Molloy, *Digitally-Disadvantaged Languages*, 11 *Internet Pol'y Rev.* art. 1654 (2022) (documenting systemic inequities facing minority-language communities in the digital sphere, including gaps in digital support and tools that damage linguistic integrity).
[36] Weihao Xuan et al., MMLU-ProX: A Multilingual Benchmark for Advanced Large Language Model Evaluation, *Proc. 2025 Conf. on Empirical Methods in Nat. Language Processing* (2025) (finding that even the best-performing models show significant cross-linguistic disparities, achieving 80.7% accuracy in English but only 57.0% in Yoruba and 58.6% in Wolof — a gap of up to 24.3 percentage points that widens dramatically for smaller models, some of which score as low as 0.6% on African languages).
[37] Cameran Ashraf, *Exploring the Impacts of Artificial Intelligence on Freedom of Religion or Belief Online*, 26 *Int'l J. Hum. Rts.* 757 (2022) (arguing that AI algorithms suppress religious speech and religious minorities' ability to practice and share their beliefs online through content prioritization and moderation systems).

*C. Viewpoint Pluralism and Democratic Deliberation*

The argument so far has concerned harms to individuals and communities. But epistemic subordination also threatens the infrastructure of democratic deliberation itself. Both the American and European traditions recognize that democratic deliberation depends on what we might call epistemic pluralism — the availability of genuinely diverse frameworks for understanding a problem. Writing on the First Amendment, Owen Fiss argues that a well-functioning public debate is one in which we can "hear voices and viewpoints that would otherwise be silenced or muffled."[38] Similarly, Cass Sunstein argues that free expression requires that "people should be exposed to materials that they would not have chosen in advance."[39] Democracy requires "unanticipated, unchosen exposures to diverse topics and ideas."[40] Interpreting Article 10 of the European Convention on Human Rights — the right to freedom of expression — the European Court of Human Rights has been equally explicit.[41] In *Centro Europa 7 v. Italy*, the Grand Chamber held that formal access to information channels is insufficient; what democracy requires is "diversity of overall programme content, reflecting as far as possible the variety of opinions encountered in the society."[42]

If democratic deliberation requires continued exposure to genuinely diverse frameworks, it is directly undermined by the cultural homogenization of model training. As we have shown, the creation of generative AI models compresses the full richness of human expression and culture into a single probabilistic model whose defaults track the median culture. Minority frameworks are not censored but absorbed, present in the training data, yet structurally subordinated in the output. It follows that in a society in which generative AI is a central knowledge infrastructure, it will to a large extent determine the baseline of democratic deliberation. Public discourse will not cease to exist, but it will increasingly take place within an epistemic field that has already been homogenized. The views developed and expressed in public debate will be constrained by the model's limited epistemic framework.

---

[38] Owen M. Fiss, The Irony of Free Speech 40–41 (Harvard Univ. Press 1996).
[39] Cass R. Sunstein, *Republic.com 2.0* 5 (2007).
[40] *Id*, 7.
[41] Convention for the Protection of Human Rights and Fundamental Freedoms art. 10, Nov. 4, 1950, E.T.S. No. 5.
[42] Centro Europa 7 S.r.l. v. Italy, App. No. 38433/09, ¶ 134 (Eur. Ct. H.R. June 7, 2012) (Grand Chamber).

The law has historically protected viewpoint pluralism through two mechanisms: safeguarding speech rights, so that dissenting and minority voices can enter public discourse; and directly regulating media, from the fairness doctrine's requirement that broadcasters present contrasting viewpoints to contemporary laws governing content moderation on digital platforms.[43] Neither remedy translates to generative AI. Speech rights ensure that speakers may speak; they do not address the epistemic infrastructure through which speech is produced, framed, and understood. Media regulation targets editorial choices made by identifiable actors. Generative AI has no editorial gate to regulate and no discrete decision to review. The legal frameworks designed to sustain viewpoint pluralism in public discourse presuppose a world in which diverse knowledge exists independently and needs only a channel through which to reach the public. Epistemic subordination collapses that assumption. The challenge for law is no longer merely to protect diverse voices and ensure they reach the public, but to preserve the diverse epistemologies and lifeworlds from which diverse speech arises.

[43]See *Red Lion Broad. Co. v. FCC*, 395 U.S. 367 (1969) (upholding the fairness doctrine requiring broadcasters to present contrasting viewpoints on matters of public importance).

## III. From Output to Infrastructure

Anti-discrimination law, cultural and linguistic rights, and laws seeking to promote viewpoint pluralism all fail to reach epistemic subordination for the same reason: they regulate downstream, while the harm is produced upstream, at the level of model training. They each seek to protect against discriminatory outcomes, safeguard minority institutions, and promote diversity of viewpoints in public discourse. But epistemic subordination is not an outcome; it is a condition embedded in the infrastructure from which outcomes emerge. The logic is straightforward: if the epistemic baseline is set during training, then training is what regulation must target.

No existing AI legal framework targets the training process in the way that has a potential to mitigate epistemic subordination. Current regulation overwhelmingly governs AI systems at the point of application. And where existing law does touch training, it does so in pursuit of other goals entirely. The EU AI Act classifies systems by the risk of their intended use, imposing transparency and conformity obligations on deployers of high-risk applications such as hiring tools, credit scoring, and law enforcement.[44] The Colorado AI Act requires developers and deployers to exercise reasonable care to avoid algorithmic discrimination in consequential decisions affecting employment, housing, insurance, and credit.[45] New York City's Local Law 144 mandates independent bias audits for automated employment decision tools before they may be used in hiring or promotion.[46] Each governs what AI systems do in specific contexts, not how they are built. The few legal instruments that do touch model training ask different questions. The EU AI Act's Article 10 requires that training datasets for high-risk systems be "relevant, sufficiently representative, and to the best extent possible, free of errors."[47] But this is a product-safety standard, not an epistemic one. The GDPR reaches training data through a different route: it restricts the processing of personal data without adequate legal basis, giving individuals rights to object to and request erasure of their data from training sets.[48] However, even a perfectly GDPR-adherent AI model will create cultural homogenization. Data protection can easily coexist with epistemic subordination.

[44]Regulation (EU) 2024/1689, of the European Parliament and of the Council of 13 June 2024 Laying Down Harmonised Rules on Artificial Intelligence (Artificial Intelligence Act), 2024 O.J. (L 1689).

The central normative claim of this essay is that protecting epistemic pluralism requires a fundamentally different regulatory orientation, one directed not at what AI systems produce but at how they are built. Because the epistemic baseline is set during the training process, any regulation that aims to preserve epistemic pluralism must intervene at that level: the composition and curation of training data, the methods by which models are aligned to human values, and the architecture of the foundation models themselves. A full regulatory proposal exceeds the scope of this short essay. But it is important to note that our argument is not merely technically speculative.

---

[45]Colorado AI Act, S.B. 24-205, 74th Gen. Assemb., 2d Reg. Sess. (Colo. 2024) (effective June 30, 2026).
[46]N.Y.C. Admin. Code §§ 20-870 to 20-874 (Local Law 144 of 2021).
[47]Regulation (EU) 2024/1689, art. 10, ¶ 3.
[48]Regulation (EU) 2016/679, of the European Parliament and of the Council of 27 April 2016 on the Protection of Natural Persons with Regard to the Processing of Personal Data (General Data Protection Regulation), 2016 O.J. (L 119) 1.

Recent machine learning research has begun to develop and propose tools that can be used to address epistemic subordination. Feng and colleagues have proposed a modular architecture in which a base model draws on a pool of smaller, community-specific language models, each trained on the knowledge and norms of a particular cultural group, preserving distinct epistemic perspectives rather than merging them into a single output.[49] Madaan and colleagues have demonstrated that culturally grounded training data can be systematically constructed for underrepresented language communities, with human reviewers from each community ensuring that cultural nuance is preserved rather than flattened through translation from dominant languages.[50] Li and colleagues have shown that existing models can be used to generate synthetic training data representing specific cultural perspectives, addressing the fundamental problem that minority-culture material does not exist at the scale required for model training.[51] And Sorensen and colleagues have redesigned the alignment process itself, training models to represent the full distribution of human values rather than collapsing diverse preferences into a single default through majority voting.[52] None of these proposals is a finished solution. But together they demonstrate that the technical capacity to intervene at the level of training is real and developing rapidly. What remains absent is the political will and the legal framework that would enable and require it.

[49]Shangbin Feng et al., *Modular Pluralism: Pluralistic Alignment via Multi-LLM Collaboration* (2024), https://arxiv.org/abs/2406.15951.
[50]Nikita Madaan et al., *Pragyaan: Designing and Curating High-Quality Cultural Post-Training Datasets for Indian Languages* (2025), https://arxiv.org/abs/2510.07000.
[51]Cheng Li et al., *CultureLLM: Incorporating Cultural Differences into Large Language Models* (2024), https://arxiv.org/abs/2402.10946.
[52]Taylor Sorensen et al., *Operationalizing Pluralistic Values in Large Language Model Alignment Reveals Trade-offs in Safety, Inclusivity, and Model Behavior* (2024), https://arxiv.org/abs/2511.14476.

## CONCLUSION

We began with a concept, epistemic subordination, and traced it through three legal domains. In each, we found the same structural gap: law that was built to protect against acts of exclusion cannot reach a technology that subordinates by inclusion, absorbing minority epistemologies into a majority default. Anti-discrimination law looks for a biased decision; epistemic subordination offers none. Cultural and linguistic rights protect institutional space; epistemic subordination operates within it. Viewpoint pluralism doctrines promote diversity of speech; epistemic subordination narrows the epistemic field in which speech is formed.

The common failure points to a common remedy. If epistemic subordination is produced at the level of model training, then law must be pushed to learn to govern at that level, seeking to shape the composition of data, the design of alignment, the architecture of the model. The technical foundations for such intervention are emerging. The political and legal foundations are not. Building them is the task ahead.